\documentclass[aas_macros]{aa}  
\usepackage{graphicx,color}
\usepackage{txfonts}
\newcommand{\chandra}{{\it Chandra}}

\newcommand{\swift}{{\it Swift}}

\newcommand{\tess}{{\it TESS}}
\newcommand{\nicer}{{\it NICER}}
\newcommand{\ms}{$M_{\odot}$}

\newcommand{\lumcgs}{ergs~s$^{-1}$}

\newcommand{\nodata}{...}
\newcommand{\nsco}{V1716~Sco}
\newcommand{\ncas}{V1405~Cas}
\newcommand{\nher}{V1674~Her}

\newcommand{\pdot}{\dot{P}}

\newcommand{\Rm}{R_{\rm m}}
\newcommand{\Rco}{R_{\rm co}}

\begin{document}

   \title{A \tess\ View of Post-Eruption Variability in  novae V1405 Cas, V1716 Sco, and V1674 Her } 


 \author{G. J. M. Luna
          \inst{1,2}
           \and
           A. Dobrotka
           \inst{3}
           \and
           M. Orio
           \inst{4,5}
                  }

   \institute{Universidad Nacional de Hurlingham (UNAHUR). Laboratorio de Investigación y Desarrollo Experimental en Computación, Av. Gdor. Vergara 2222, Villa Tesei, Buenos Aires, Argentina\\
   \email{juan.luna@unahur.edu.ar}
   \and
   Consejo Nacional de Investigaciones Científicas y Técnicas (CONICET), Argentina.
    \and
Advanced Technologies Research Institute, Faculty of Materials Science and Technology in Trnava, Slovak University of Technology in Bratislava, Bottova 25, 917 24 Trnava, Slovakia
\and
    INAF-Osservatorio Astronomico di Padova, vicolo Osservatorio, 5,
35122 Padova, Italy
\and
Department of Astronomy, University of Wisconsin, 475 N. Charter Str., Madison WI 53706, WI, USA
}

   \date{Received ; accepted }

\abstract
{We analyzed \tess\ archival data of three novae after recent outbursts, searching the orbital and white dwarf (WD) rotation period and possible variations of these periods. In \ncas, we detected a period of $\sim$116.88 seconds, which we identified as due to the WD spin, and measured a rate of increase of 0.00165$\pm0.000006\, {\rm s\, d}^{-1}$, one of the fastest spin-down rates ever recorded. The rapid spin-down coupled with an X-ray luminosity several orders of magnitude lower than the available spin-down power, strongly indicates that the system is in a magnetic ``propeller'' state, namely the rotational energy powers the system's X-ray luminosity. We measured a previously unknown orbital period of 1.357$\pm0.005\,{\rm days}$ for \nsco. If the X-ray flux modulation with a period of 77.9 s detected in outburst for this nova is due to the rotation of an strongly magnetized white dwarf as in other novae with similar modulations of the supersoft X-ray source in outburst, the system is in a parameter space that challenges standard models of cataclysmic variable evolution. For \nher, which has already been classified as an intermediate polar (IP), we confirm the known spin period of 501.328$\pm0.024\,{\rm s}$ and the orbital period of $0.15293 \pm 0.00004$ days, suggesting that the spin modulation was also the root cause of the periodicity in X-rays in outburst, and that the WD atmosphere in the supersoft X-ray phase was not thermally homogeneous. Our results highlight the power of high-cadence, continuous observations in revealing extreme and unexpected characteristics of accreting white dwarfs.}


   \maketitle
%

\section{Introduction}

Classical novae are the manifestation of thermonuclear bursts on the surfaces of white dwarf stars in interacting binary systems \citep[e.g.,][]{2008clno.book.....B,2020ApJ...895...70S}. Although novae are now established as multi-wavelength and possibly even multimessenger sources \citep[][]{2025arXiv250707096T}, the study of the optical light curves still provides a very important window into the complex physics of accretion, thermonuclear runaway, and the subsequent ejection of matter. However, our understanding of these cataclysmic events has historically been limited by the constraints of ground-based observations. 
    
The advent of the Transiting Exoplanet Survey Satellite \citep[\tess;][]{2015JATIS...1a4003R} provides an unprecedented opportunity to study novae in exquisite detail. \tess\ observations have led to the discovery of numerous new orbital periods for classical novae, either in archival observations done before \citep{v462Orbit,v572Orbit} the nova discovery, or after the eruption \citep[][on \nher]{2024BAAA...65...60L}. Furthermore, the high-precision, continuous light curves from \tess\ have revealed a complex tapestry of variability during nova eruptions that was previously undetectable \citep[e.g.][]{Sokolovsky2023}. These discoveries provide crucial new constraints for theoretical models of nova physics. 

Novae occur also in intermediate polar (IP) systems. In IPs, the white dwarf’s magnetic field is strong enough to disrupt the inner accretion disk and funnel material onto the magnetic
poles \citep[e.g.][]{1995CAS....28.....W}. In some novae, the WD spin period was measured already in outburst, during the super-soft phase, through observations obtained with X-ray satellites \citep[e.g.][and references therein]{2025Univ...11..105B}.

This paper builds upon the growing body of work that uses the unique capabilities of \tess\ to explore the multifaceted nature of classical novae. We specifically address three novae that may host strongly magnetized WDs.  We analyze \tess\ light curves of  \ncas, \nsco\ and \nher, obtained before and/or after their outbursts. Section 2 explains which data and methods we have used for this project. Section \ref{sec:results} outlines our discoveries regarding each nova. Section \ref{sec:disc} contains the discussion of our findings, and Section \ref{sec:conc} presents our conclusions.

\section{\tess\ data \label{sec:obs}}

Table \ref{tab:1} presents the Sectors and the starting dates of the observations conducted with \tess. Only in the case of \ncas, we found pipeline-ready light curves at the Barbara A. Mikulski Archive for Space Telescopes (MAST)\footnote{\url{https://archive.stsci.edu }} with a 20 s cadence (Prop. ID G06168, PI: Tzanidakis, A.). For \nsco\ and \nher, we used the Python script \texttt{quaver} to extract the background-corrected light curves from the target pixel file (TPF) created from a cut on the Full Frame Images (FFIs) \citep{tesscut} and then selected a region of 2$\times$2 pixels centered on the source's SIMBAD coordinates. Although \texttt{quaver} was mainly designed to study the stochastic variability of AGNs observed with \tess, where an accurate modeling of the background is necessary, \texttt{quaver} provides a quick-and-easy way to obtain background-corrected light curves on which periodicities can be searched. We refer the reader to \citet{quaver} for further details about how the background is modeled.

We used the generalized Lomb-Scargle (GLS) algorithm as implemented in the \texttt{astropy} \citep{Astropy2022} library to search for periods in the power spectrum. We adopted the standard normalization provided by the \texttt{astropy} implementation. Following the general principles outlined in \cite{Vaughan2005} and \cite{Feigelson2018} for statistically robust period searches in the presence of red-noise continuum (flickering), we define significance at the 99.99\% using a frequency-dependent threshold derived from an Autoregressive (AR) model \citep{Box2015}. We model the noise continuum of the power spectrum using an AR process of order $p$~[AR($p$)], where the order $p$ is determined by the Akaike Information Criterion \citep[AIC][]{Akaike1974} to capture the correlation timescales of the flickering.
Based on this model, we derived a frequency-dependent significance threshold (FDST). To determine the uncertainties in the measured periodicities, we employed a Monte Carlo (MC) simulation integrated with our AR noise modeling. We first characterized the local red noise (flickering) and then generated 200 synthetic light curve realizations. Each realization consisted of a sinusoidal signal —- with an amplitude and period matching the best-fit values from the observed data —- injected into a synthetic noise background. Unlike standard white-noise simulations, our noise background was generated by passing Gaussian white noise through a linear filter defined by the specific AR coefficients derived from that window's residuals. This ensures that the synthetic data preserves the "colored" noise profile (flickering). The $1\sigma$ period uncertainty was then defined as the standard deviation of the resulting Lomb-Scargle peak positions across the 200 realizations. These uncertainties were subsequently used as weights in a Weighted Least Squares (WLS) fit to study the secular period change rate ($\dot{P}$).

\renewcommand{\arraystretch}{1.3}
\begin{table*}
\small
\caption{\label{tab:1} \tess\ observations log. We list the orbital and spin periods determined in this work. An error is not provided for the spin period of \ncas\ given its rapid change during the course of the observations.}
\centering
\begin{tabular}{lcccccc}
\hline\hline
Nova & Sectors & Start date  & Eruption date & P$_{spin}$ [s] & P$_{orb}$ [days ] & Days since outburst \\
\hline
V1716~Sco & 12 & 2019 May 21 & 2023 Apr 20 & \nodata &  $1.3665 \pm 0.0099$  & -1430 \\
 & 39 & 2021 May 27 &  & \nodata & $1.3571 \pm 0.0101$  & -693\\
 & 66 & 2023 Jun 02 &  & \nodata & \nodata & 43  \\
 & 93 & 2025 Jun 03 &  & \nodata & $1.3528 \pm 0.0067$  & 775 \\
V1405~Cas & 77 & 2024 Mar 26 & 2021 Mar 18  & 116.88 & $0.188395 \pm 0.000008$ & 1103 \\
            & 78 & 2024 Apr 23 & & \nodata & & 1132 \\
V1674~Her & 40 & 2021 Jun 24 &  2021 Jun 12  & & $0.15274 \pm 0.00029$ & 12\tablefootmark{2} \\
    & 53 & 2022 Jun 13 & & \nodata & $0.15288 \pm 0.00010$  &  366  \\
    & 54 & 2022 Jul 09 & & \nodata & $0.15298 \pm 0.00006$ & 392 \\
    & 80 & 2024 Jun 18 & & $501.328 \pm 0.024$ & $0.15289 \pm 0.00193$ & 1102\\
\hline
\tablefoottext{2}{from \citep{2024BAAA...65...60L}}
\end{tabular}
\end{table*}

\section{The novae and our new results \label{sec:results}}

\subsection{\ncas\ (Nova Cas 2021)}


Yuji Nakamura identified a nova eruption on \ncas\ registering a magnitude of 9.6 (unfiltered) on March 18 2021 at 10:10 UT. The nova has been followed with Neil Gehrels $Swift$ Observatory since the outburst, and recent observations in 2024 and 2025 still detect a supersoft X-ray source (SSS), albeit at lower luminosity than most SSS in novae \citep[see][]{Page2024,woodward2025,DiGiacomo2025}. The AAVSO (American Association of Variable Star Observers) light curve showed a rise to maximum within days, followed by a long period of stunted behavior with multiple peaks around maximum magnitude up to V$\approx$ 5, becoming at times visible to the naked eye \citep{2021ATel14614....1M}. The nova in 2025 is back at minimum optical light. Recent optical spectra still show many emission lines, yet the spectrum is no longer typical of the late-phase outburst of novae \citep{woodward2025}.
 
\begin{figure}
\begin{center}
\includegraphics[scale=0.38]{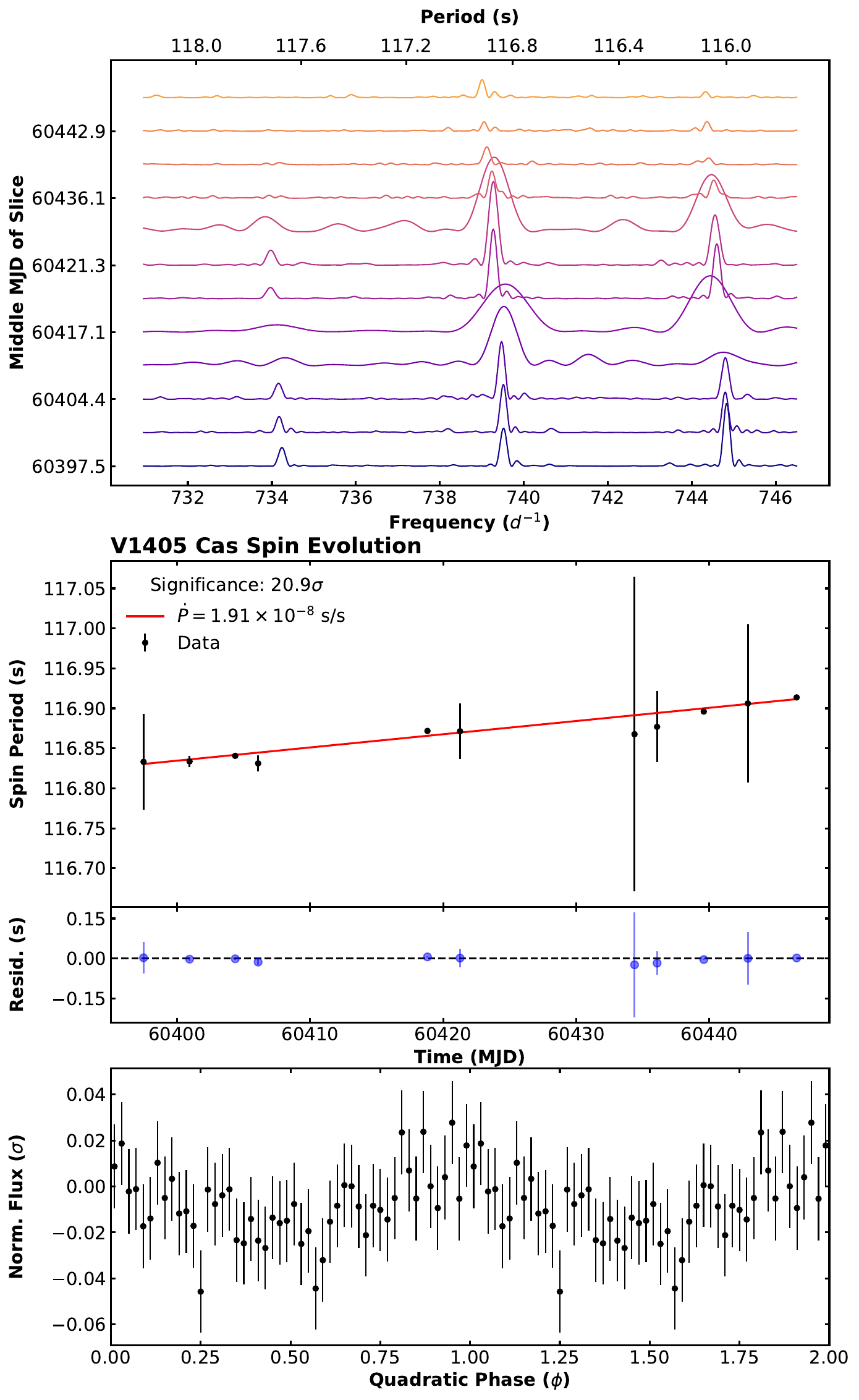}
\caption{{\it Top}: Waterfall plot of stacked Lomb-Scargle periodograms. The right Y-axis indicates the Middle MJD of each 5-day slice, with the color gradient representing temporal progression. It illustrates the secular evolution of the spin frequency over the $\sim$56-day \tess\ observation baseline. Flanking the central period, two periods are observed that we identify as the beat periods between the orbital and spin periods (see Section 3.1). {\it Middle}: Evolution of the spin period as a function of time (MJD). The black points denote measurements from individual sliding windows, with 1$\sigma$ uncertainties derived from 200-iteration Monte Carlo simulations integrated with an $AR(p)$ red noise model. The red line represents a Weighted Least Squares (WLS) linear fit, yielding a secular spin-down rate of $\dot{P}$=1.91$\times$10$^{-8}\pm$7.01$\times$10$^{-10}$ s/s, representing a 20.9$\sigma$ significance level over the constant hypothesis. The residuals of the linear period fit, show no systematic trends or higher-order derivatives. {\it Bottom}: Pulse profile (50 bins) constructed via quadratic phase folding using the derived ephemeris ($\phi(t) = f_0 \Delta t + \frac{1}{2} \dot{f} \Delta t^2$). The phase-folded profile is dominated by a broad, quasi-sinusoidal modulation. The successful recovery of this complex structure across the full \tess\ baseline confirms high phase coherence ($Q=1/|\dot{P}| \approx 5.2 \times 10^7$) and validates the quadratic timing correction ($\dot{P} = 1.91 \times 10^{-8}$ s/s).}
\label{fig:psdV1405}
\end{center}
\end{figure}

We note that \citet{2021RNAAS...5..150S} documented a period of 0.1883907$\pm$0.0000048 days based on \tess\ data from 2019 and 2020. In 2024, across Sectors 77 and 78, \tess\ recorded light curves with a 20-second cadence over 56.8 days for \ncas. Our periodogram shows a significant peak at the frequency equivalent to the previously reported orbital period, P$_{orb}$=0.188395$\pm$0.000008 days. At considerably higher frequencies, which can now be resolved with the 20-second interval data, a prominent peak appears at 739.3\, {\rm d}$^{-1}$. However, as observed in the symbiotic X-ray binary GX~1+4 \citep{2023A&A...676L...2L}, additional peaks were detected near the main one, suggesting the period might have shifted during the observation. Consequently, we derived periodograms from five-day consecutive and overlapping (50\% overlap) sections of the light curve. Successive slices indicated the period was decreasing at a drift rate of $0.00165 \pm 0.00006\, {\rm s\, d}^{-1}$ (Fig. \ref{fig:psdV1405}). Notably, two additional peaks were apparent flanking the 739.3 \, {\rm d}$^{-1}$ peak. We interpret these peaks as beat periods of $\omega$-$\Omega$ = 734.17864\, {\rm d}$^{-1}$ and $\omega$+$\Omega$ = 744.7948\, {\rm d}$^{-1}$ between the spin and orbital periods, with the central peak at a frequency $\omega$ potentially representing the white dwarf's spin period while $\Omega$ represents the frequency of the orbital period. The possibility that the peak at the highest frequency, 744.7948\, {\rm d}$^{-1}$, is the spin period remains open in view of the models presented by \cite{Ferrario1999}. In this instance, the central peak correlates with $\omega$-$\Omega$ = 739.4867643\, {\rm d}$^{-1}$, while the leftward peak is associated with $\omega$-$2\Omega$ = 734.17864\, {\rm d}$^{-1}$. High signal-to-noise X-ray timing measurements are essential to distinguish these. Nonetheless, our primary discovery regarding period drift and its potential consequences holds true (refer to Section \ref{sec:conc}).


\subsection{\nsco\ (Nova Sco 2023)}


\nsco\ was detected as a classical nova on 2023 April 20 (discovered by Andrew Pearce on 2023 April 20.678 UT). In the super-soft X-ray phase, the X-ray light curves obtained with \nicer\ \citep{Nicer2016} and \chandra\ \citep{Chandra2005} revealed a periodicity of 77.9 seconds \citep{2023ATel16167....1D,Worley2025}. From the detection of this period, it has been debated whether \nsco\ belongs to the intermediate polar class of magnetic cataclysmic variables.

\begin{figure*}
\begin{center}
\includegraphics[scale=0.33]{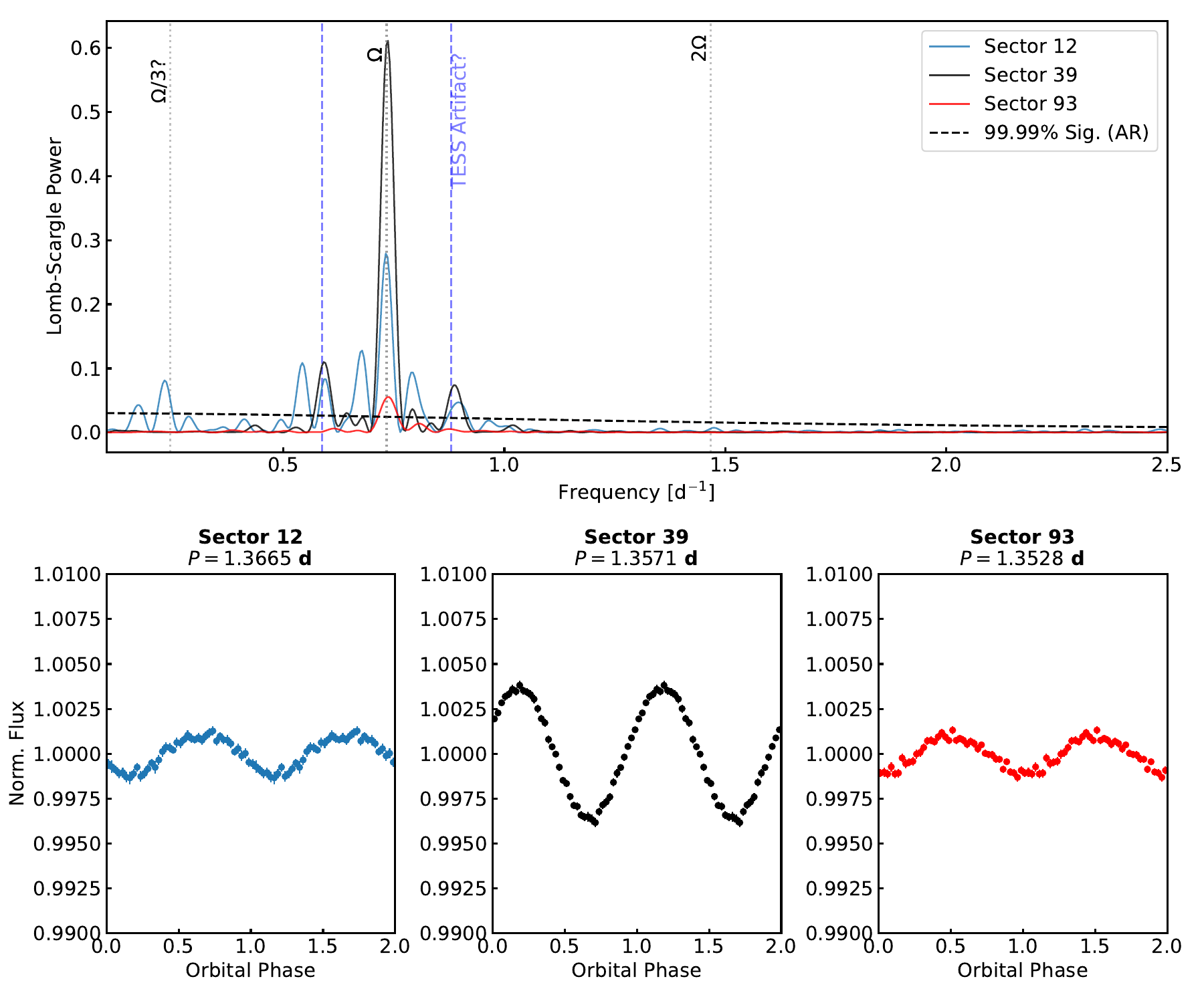}
\includegraphics[scale=0.32]{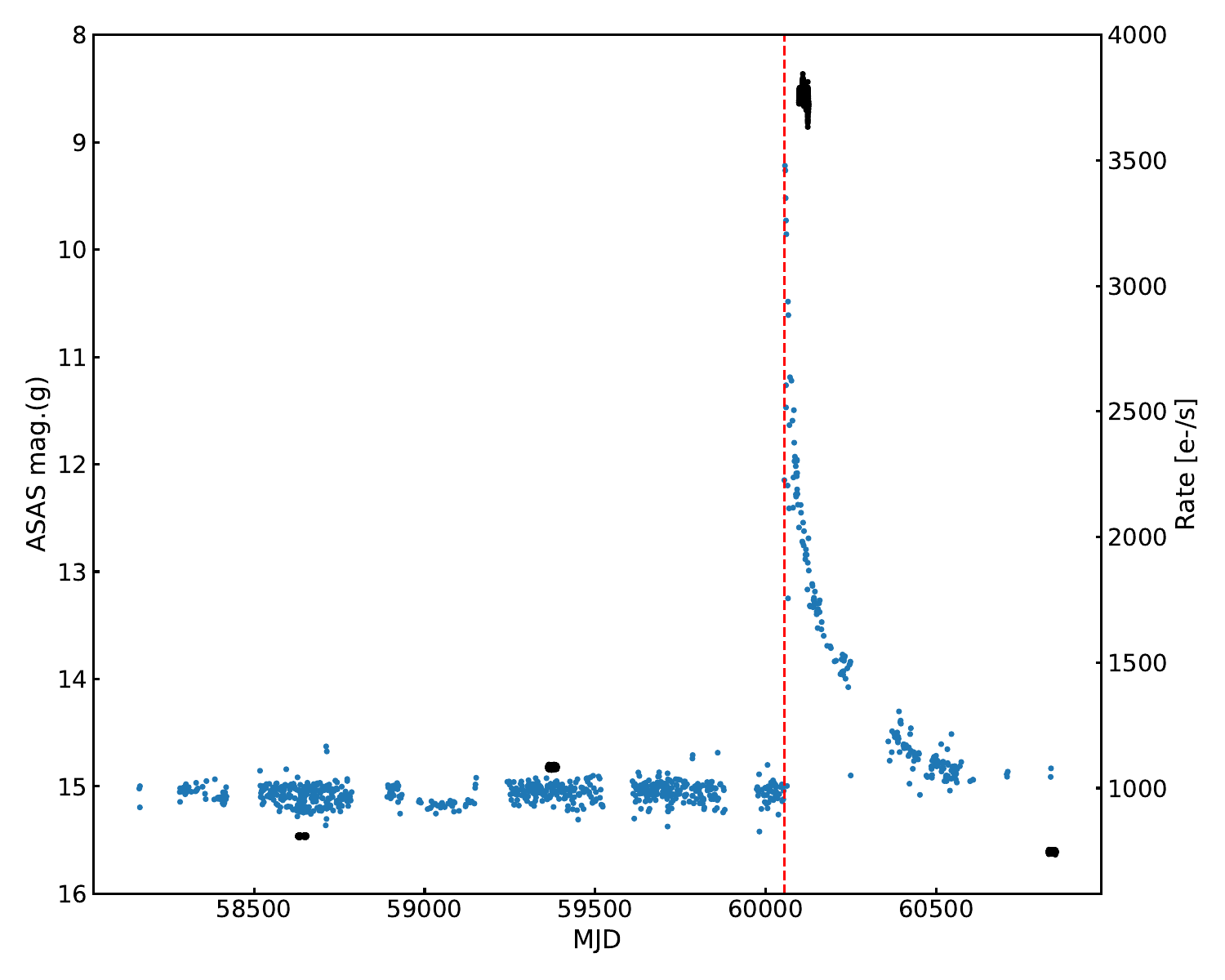}
\caption{(Top): Lomb-Scargle periodograms for \nsco\ across Sectors 12, 39, and 93. The primary peak at $\sim 0.735 \text{ d}^{-1}$ corresponds to the orbital frequency ($\Omega$), with the first harmonic ($2\Omega$) also visible at $\sim 1.47 \text{ d}^{-1}$. The dashed blue vertical lines indicate potential instrumental artifacts related to the \tess\ orbital period ($\sim 13.7$ d) and its harmonics, which appear as sidebands to the main signal ($\Omega-2\Omega_{TESS}$; $\Omega+2\Omega_{TESS}$). The curved dashed black line represents the 99.99\% significance threshold derived from an Autoregressive [AR($p$)] noise model. This frequency-dependent baseline accounts for the red-noise continuum (flickering) inherent in the system, confirming that the orbital modulation remains statistically significant despite the increased noise power at low frequencies. {\it Bottom left:} Phase-folded light curve of Sectors 12, 39 and 93 at their detected orbital period, binned into 50 phase intervals to reduce point density and highlight the orbital modulation. {\it Right}: ASAS-SN $g$-band magnitude (blue points, left Y-axis) light curve of \nsco\ before, during and after the nova outburst recorded on MJD=60055.178. The black points (right axis) show the raw \tess\ instrumental count rate. Note that the \tess\ Y-axis is independent and count rates are not scaled to the ASAS-SN $g$-band magnitude; sector-to-sector jumps in the \tess\ rate are due to varying background and crowding levels and do not affect the timing analysis.}
\label{fig:psdV1716}
\end{center}
\end{figure*}

\nsco\ was observed during \tess\ Sectors 12, 39, 66 and 93. The periodograms of sectors 12, 39 and 93 show a strong peak at the frequencies corresponding to a period of approximately 1.36 days (see Table \ref{tab:1}). Figure \ref{fig:psdV1716} shows the power spectra with the FDTS at the 99.99\% confidence level and and each sector' light curve folded at their best period with T$_{0}$ as that from the first point in the Sector 12 light curve.
The period is not detected on the periodogram corresponding to Sector 66 which started 43 days after the nova outburst and covered part of the fading phase after maximum, as illustrated in Figure \ref{fig:psdV1716} where we show the \tess~observations on top of the ASAS-SN light curve \citep{ASAS1,ASAS2}. We analyzed the stability of the orbital period across Sectors 12, 39, and 93. The independent period measurements are $P_{S12} = 1.3665 \pm 0.0099$ d, $P_{S39} = 1.3571 \pm 0.0101$ d, and $P_{S93} = 1.3528 \pm 0.0067$ d. A constant-period fit to these data yields a weighted mean of $\bar{P}_{orb} = 1.357 \pm 0.005$ d with a reduced $\chi^2$ of 0.66. This indicates that the orbital period is stable within the measurement precision over the $\sim 6$-year baseline. The sinusoidal pattern observed in the \tess\ light curves leads us to identify it as the orbital period. 

The supersoft X-ray source (SSS) phase in outburst had an X-ray flux modulation with a definite period of 77.9 seconds \citep{Dethero2023,Wang2024,Worley2025}. \citet{Worley2025} found that the irregular period drift measured by \cite{Wang2024} was due to an incorrect estimate of the statistical uncertainty in the measurement of a periodicity with variable amplitude, so this period was most likely stable. These authors suggested that it is due to the WD spin, rather than a stellar pulsation. A SSS modulation due to rotation, rather than a stellar pulsation, seems to indicate a WD atmosphere with inhomogeneous temperature, presumably higher at the polar caps, which is still unexplained and difficult to model. However, there are several other observations showing rotational modulation of the SSS flux in novae. This type of periodicity (albeit usually longer, of several minutes) in the super-luminous SSS was detected again in quiescence in the accreting phase in four other novae, namely V4743~Sgr \citep[][and references therein]{Zemko2016,Dobrotka2017}, V2491~Cyg \citep{Zemko2015}, V407~Lup \citep{Aydi2018,Orio2024} and in \nher\ \citep[][and this paper]{Drake2022}, and all of these novae have now been classified as IPs. If this is also the case of \nsco \, this nova has an uncommonly long orbital period and an exceptional P$_{spin}$/P$_{orb}$ ratio of 0.00066 (refer to Sect. \ref{sec:conc}).

\subsection{\nher\ (Nova Her 2021)}

\begin{figure*}
\begin{center}
\includegraphics[scale=0.4]{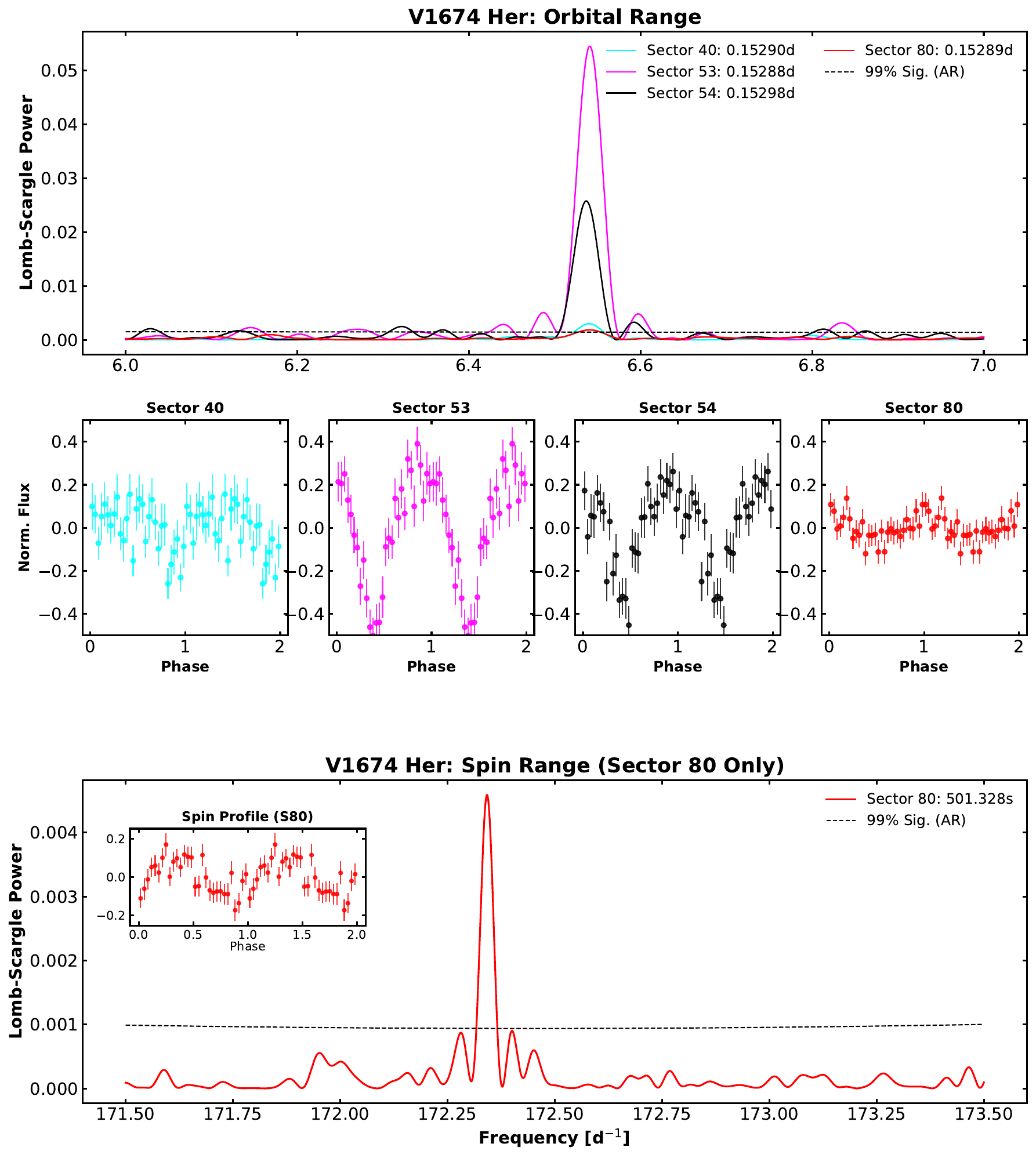}
\caption{Comprehensive timing analysis and evolution of the periodic signals in \nher. {\it Top}: Lomb-Scargle power spectrum in the orbital frequency range ($6.0$–$7.0 \text{ d}^{-1}$) for Sectors 40 (cyan), 53 (magenta), 54 (black), and 80 (red). The dashed line represents the $99\%$ significance threshold for Sector 80, calculated using an AR($p$) red noise model to account for stochastic flickering. {\it Middle}: Individual phase-folded orbital profiles for each sector, taking T$_0$=59873.817 from \citet{v1674_ephem} and binned into 30 phase intervals with $1\sigma$ error bars. {\it Bottom}: Lomb-Scargle power spectrum in the spin frequency range ($171.5$–$173.5 \text{ d}^{-1}$) from Sector 80. The inset shows the binned spin pulse profile with associated $1\sigma$ uncertainties. All reported periods and uncertainties were derived using 200-iteration AR-integrated Monte Carlo simulations to ensure robustness against the colored noise background of the post-nova state.}
\label{fig:psdV1674}
\end{center}
\end{figure*}

\nher\ was discovered in outburst on 2021 June 12.537 UT at 8.4 magnitudes by Seiji Ueda
(Kushiro, Hokkaido, Japan) reaching naked-eye magnitudes at its peak. A period of 501.42~s was reported by \citet{2021ATel14720....1M} found in ZTF (Zwicky Transient Facility) data obtained during quiescence. Close periodicities have since been detected in X-ray data by \citet[501.42 s;][]{2021ApJ...922L..42D}, \citet[501.535; ][]{2022ApJ...932...45O}, and
\citet[between 501.4–501.5 s; ][]{2024MNRAS.528...28B}. The orbital period of 0.15290$\pm$0.00003 days was later reported by \citet{2021ATel14835....1S} and found in \tess\ Sector 40 data taken 12 days after the eruption by \citet[][0.1529$\pm$0.0001 days]{2024BAAA...65...60L}. The study of Sectors 40, 53, 54 and 80 has revealed the presence of the orbital period in all sectors albeit with a lower significance during Sector 80 as shown by the FDST level in the top panel of Fig. \ref{fig:psdV1674}. In the power spectrum from Sector 80, the spin period, identified as P$_{spin}$=$501.328 \pm 0.024$ s, is observed. We conducted a multi-sector timing analysis to track the evolution of the orbital signal, with no detectable changes on the period but an apparent evolution on the amplitude. However, given that the \tess\ data are not flux-calibrated, these changes in amplitude are difficult to isolate from variable sector-to-sector background.

\section{Discussion \label{sec:disc}}

\subsection{The Unprecedented Spin-Down of \ncas}

The GLS periodogram of \ncas\ reveals a dominant periodicity at $116.88$\,s, which we interpret as the WD spin period. By dividing the \tess\ light curve into sequential segments and measuring the period in each, we detected a secular increase in the spin period, corresponding to a spin-down rate of $\pdot = 1.91 \times 10^{-8}$\,s\,s$^{-1}$ (0.00164$\, {\rm s\, d}^{-1}$). This rate is among the highest ever measured in WD binaries, even for IPs (see Table 2). This allows us to derive a fundamental property of the system within the magnetic rotator framework, i.e. the spin-down luminosity, $L_{\text{sd}} = 4\pi^2 I \dot{P}P^{-3} \approx 4.72 \times 10^{36}$ \lumcgs \citep{Shapiro1983}\footnote{Assuming a WD mass of 0.7\ms\ \citep{Taguchi2023} in the moment of inertia \citep[$I \approx kM_{WD} R_{WD}^{2} \approx 10^{49}$ g cm$^2$, with $k$=0.12 from][]{Yoon2005}}. This immense power release is sufficient to dominate the system's energy budget.

A braking torque of this magnitude is inconsistent with standard disk-fed accretion, which is expected to transfer angular momentum to the WD, causing spin-up. We therefore propose that \ncas\ is in a magnetic propeller state \citep{Illarionov1975}. In this regime, the magnetospheric radius ($\Rm$) exceeds the co-rotation radius ($\Rco$), leading to the centrifugal ejection of matter attempting to accrete, which extracts angular momentum from the WD. 

The detection of beat frequencies in the power spectrum is often considered a signature of disk-fed accretion, though we note that such frequencies can also arise from the periodic illumination of the companion star in the optical regime \citep{Wynn1992}. This adds a layer of interpretative complexity, as a detailed analysis reveals an evolving beat frequency structure. While the classic \(\omega -\Omega \) orbital sideband is present at times, we also detect a prominent \(\omega +\Omega \) sideband, which suggests a more complex, non-linear accretion geometry. The spin-down rate we measure for the white dwarf is physically difficult to reconcile with a steady-state disk. Instead, we propose that \ncas\ is in a disrupted, discless propeller state. In this regime, the accretion flow might consists of discrete, diamagnetic 'blobs' that interact directly with the rapidly rotating magnetosphere \citep{Wynn1997}. The presence of both sidebands in \ncas\ is consistent with the amplitude modulation of the spin pulse \citep{Warner1986}. It should be noted that while \citet{Warner1986} originally interpreted such sidebands as the result of modulation within a steady-state disk, the extreme spin-down rate measured here suggests that this modulation instead arises from the disrupted, high-torque environment of the propeller state. In this regime, the accretion geometry is dominated by the propeller mechanism rather than a symmetric disk. Such a mechanism may provide an efficient mean for angular momentum transfer, potentially contributing to the observed spin-down torque.

While the propeller model provides a theoretical framework for the observed spin-down, the required mass-ejection rate of $\dot{M} \approx 8 \times 10^{20} \text{ g s}^{-1}$ is not supported by observations. In the propeller regime, the rotational energy loss must be balanced by the power required to gravitationally unbind and eject matter from the system. The power needed to eject a mass rate $\dot{M}_{\text{eject}}$ from the magnetospheric radius ($R_m$) is approximately $L_{\text{eject}} \approx G M_{\text{wd}} \dot{M}_{\text{eject}} / R_m$. By setting $L_{\text{eject}}$ equal to the spin-down luminosity ($L_{\text{sd}} \approx 4.72 \times 10^{36}$ erg s$^{-1}$) and assuming $R_m$ is on the order of the co-rotation radius ($R_{\text{co}} \approx 3.2 \times 10^9$ cm for a 0.7 M$_{\odot}$ white dwarf, \citep{Nauenberg1972}, we can estimate the required mass ejection rate of  $\approx$ 1.51$\times$ 10$^{20}$ g s$^{-1}$. Recent spectroscopic analysis by \cite{DiGiacomo2025} suggests that the late-time emission in \ncas\ is dominated by ballistically expanding ejecta from the 2021 eruption, without the need for a persistent accretion disk. The absence of a massive outflow or extended emission consistent with such high $\dot{M}$ values suggests that the extreme rotational energy loss ($L_{sd} \approx 10^{36} \text{ erg s}^{-1}$) is likely dissipated through non-thermal channels, such as magnetic dipole radiation, effectively making \ncas\ a high-power white dwarf pulsar candidate.

The spin evolution of \ncas\ differs from that of the other recent fast nova studied in this paper, \nher. Immediately following its outburst, \nher\ experienced a brief but powerful spin-down ($\dot{P} \approx$ +5.4$\times$10$^{-8}$ s s$^{-1}$), which was attributed to angular momentum conservation in the expanding nova shell \citep{2022ApJ...940L..56P}. However, this effect was transient. After just 15 days, the system transitioned to a state of rapid, accretion-driven spin-up at a rate of $\dot{P} \approx$ -5.8$\times$10$^{-9}$ s s$^{-1}$. The persistent, powerful spin-down we measure in \ncas\ seems to be fundamentally different. It lasted at least $\sim$56 days, suggesting a sustained braking torque, consistent with a magnetic propeller state. This comparison highlights that while multiple mechanisms can affect the white dwarf's spin in the aftermath of a nova, the enduring and strong spin-down in \ncas\ points squarely towards a propeller-dominated regime.
\subsubsection{The Missing X-ray Pulsation}

The 116.88 s spin period, while prominent in the optical, remains undetected in current \swift\ X-ray observations. Should future high signal-to-noise observations confirm the absence of this signal, it would support the propeller scenario. In this framework, the thermal X-ray emission from any existing accretion columns is likely suppressed by a high optical depth environment at the disk-magnetosphere boundary. This opacity is a direct consequence of the high spin-down luminosity, which drives significant energy dissipation and turbulence. This regime differs from lower-luminosity systems like AR~Sco, where a more transparent environment allows for the detection of non-thermal, pulsed X-ray emission (Takata et al. 2018). Consequently, the lack of X-ray modulation in \ncas\ may serve as a diagnostic of the extreme physical conditions and high mass-loading characteristic of its current propeller state.

\subsubsection{Context Among Rotation-Powered Systems}
The timing properties, and their implications for the energy budget of \ncas\ are extreme when compared to other known systems (Table \ref{tab:comparison}). Its spin-down rate is over five orders of magnitude greater than that of other well-established WD propellers such as AE~Aqr and LAMOST~J024048.51+195226.9. 
%
\begin{table*}[ht]
\centering
\caption{Comparative Properties of Spinning-down white dwarfs.}
\label{tab:comparison}
\begin{tabular}{lccccc}
\hline
\hline
Property & V1405 Cas & AR Sco & LAMOST J0240 & AE Aqr &  Gaia22ayj \\
\hline
P (s) & 116.88 & $\sim$118 & 24.93 & 33.08 & 561.6 \\
$\dot{P}$ (s/s) & $\approx 1.9 \times 10^{-8}$ & $\approx 3.9 \times 10^{-13}$ & $\approx 2.8 \times 10^{-14}$ & $\approx 5.6 \times 10^{-14}$ & $\approx 2.9 \times 10^{-12}$ \\
$L_{sd}$ (erg/s) & $\approx 4.7 \times 10^{36}$ & $\approx 1.5 \times 10^{33}$ & $\approx 1.2 \times 10^{33}$ & $\approx 6 \times 10^{33}$ & $\approx 6.5 \times 10^{31}$ \\
\hline
\end{tabular}
\tablefoot{References for comparison systems: AE Aqr \citep{Wynn1997}; LAMOST J0240 \citep{Jiang2024}; Gaia22ayj \citep{rodriguez2025link}; AR Sco \citep{Marsh2016}. In the case of AR~Sco, its spin-down has been a matter of debate, see \cite[][and references therein]{Pelisoli2022}.}
\end{table*}

A critical, testable prediction of this model is the generation of a distinct thermal emission component. The immense energy dissipated at the turbulent disk-magnetosphere boundary can heat the region to produce a super-soft X-ray signature \citep{Erkut2019}. The detection of such a component in \ncas\ provides supporting evidence for the propeller mechanism. This very soft component, which we previously referred to in our X-ray luminosity calculations, is also noted by \citet{DiGiacomo2025}. Their findings indicate a rise in luminosity within the 0.3-1.0 keV energy range, alongside a decline in the temperature of the blackbody component about 1350 days after the outburst, suggesting an enlargement of the emitting region's area.

\subsubsection{\ncas\ as an extreme spin-down powered system}

The discovery of a rapid \mbox{116.88-s} spin period and a \mbox{4.52-hr} orbital period in \ncas\ places it in a remarkably similar parameter space to the archetypal spin-down powered white dwarf propeller, AR~Sco, which exhibits a spin period of \mbox{$\sim$118.2-s} and an orbital period of \mbox{$\sim$3.56-hr} \citep{Marsh2016}. This striking resemblance in fundamental timescales suggests a common underlying physical nature.

However, despite these similarities, the rotational energetics of \ncas\ are remarkably different. Our calculated spin-down luminosity of \mbox{$L_{\rm sd} \approx 4.72 \times 10^{36}$ erg\,s$^{-1}$} is more than four orders of magnitude greater than the \mbox{$L_{\rm sd} \approx 1.5 \times 10^{33}$ erg\,s$^{-1}$} derived for AR~Sco \citep{Marsh2016}. This power output is driven by an extreme spin-down rate of \mbox{$\dot{P} \approx 1.91 \times 10^{-8}$ s\,s$^{-1}$}, which itself is five orders of magnitude larger than that of AR~Sco \citep[$\dot{P} \approx 3.9 \times 10^{-13}$ s\,s$^{-1}$;][]{Marsh2016}.

This large disparity suggests an evolutionary link, positioning \ncas\ as a nascent, exceptionally powerful propeller system, perhaps recently spun-up or re-energized by the 2021 nova eruption. While AR~Sco represents a more mature, stable state of magnetic propeller evolution, \ncas\ gives us an unprecedented glimpse into the earliest and most energetic phase of this phenomenon. Over a secular timescale, as its rotational energy reservoir is dissipated, \ncas\ may evolve into a system akin of AR~Sco.

\subsubsection{Are the 116.88-s period and its fast evolution due to dwarf novae oscillations (DNOs)?}
Dwarf Nova Oscillations (DNOs) are quasi-coherent modulations typically attributed to the rotation of a low-inertia accretion layer at the white dwarf equator, characterized by period instability and a strong dependence on the instantaneous accretion rate \citep{Warner2004}. However, the 116.88-s signal in \ncas\ observed in \tess\ Sectors 77 and 78 exhibits a coherence ($Q=|1/\dot{P}| \approx 5.2 \times 10^7$) that exceeds the typical DNO coherence ($Q \approx 10^3 - 10^6$) by over an order of magnitude.

Furthermore, the interpretation of this signal as a DNO is challenged by the system's late-time behavior. AAVSO monitoring and X-ray observations show no evidence of dwarf nova-like outbursts or high-accretion states during the 2024 \tess\ window \citep[see e.g.][]{DiGiacomo2025}. DNOs are fundamentally transient and generally disappear as $\dot{M}$ declines; the persistence of such a highly coherent signal in a quiescent environment favors a solid-body white dwarf spin interpretation.

\subsection{The Evolutionary Enigma of \nsco}

If in \nsco\ the 77.9 s period is due to the spin and a stellar pulsation is ruled out - like the constancy of the period found by \citet{Worley2025} appears to imply, such a short rotational period with the long 1.357$\pm$0.005-day orbital period represents a significant challenge to conventional models of CV evolution. The system would occupy a peculiar place in the $P_{\text{spin}}$-$P_{\text{orb}}$ diagram\footnote{listed by K. Mukai at \url{https://asd.gsfc.nasa.gov/Koji.Mukai/iphome/iphome.html}}, shared by only a handful of other objects (see Fig. \ref{fig:pspin_porb}), requiring a re-evaluation of the physical processes that govern the formation and evolution of magnetic CVs. Its wide orbit, in particular, is difficult to reconcile with standard evolutionary pathways and could be explained by two distinct scenarios.

The first possibility is that the binary passed through a common envelope (CE) phase that was unusually inefficient at removing orbital angular momentum. During the CE phase in close binary systems, the giant progenitor of the white dwarf engulfs its companion. Because of the complex physics, in most models the result of the common envelope phase is parametrized with the so-called CE efficiency parameter, $\alpha_{\text{CE}}$, which describes how effectively orbital energy is converted into kinetic energy to expel the envelope \citep{Webbink1984}. The final orbital separation is very sensitive to this parameter, because a low value of $\alpha_{\text{CE}}$ implies that a larger amount of orbital energy must be used to successfully unbind the envelope, leading to a much tighter final orbit. The wide 1.36-day orbit of \nsco\ would therefore require a high efficiency in converting the orbital into kinetic, otherwise there must have been an alternative energy source, such as recombination energy \citep{Ivanova2013}. Thus, \nsco\ may be an important observational benchmark, showing that some systems may evolve through pathways that avoid the dramatic orbital shrinkage typically associated with the CE phase. 
The evolution of such systems is highly sensitive to the common envelope (CE) efficiency parameter, $\alpha_{CE}$. While the exact value of $\alpha_{CE}$ remains a subject of active debate \citep{Ivanova2013, Webbink2008}, several population synthesis studies have invoked high efficiencies ($\alpha_{CE} \approx 1$) to reconcile theoretical birthrates with the observed populations of post-CE binaries \citep{Davis2010, Toonen2013}. This debate often centers on the inclusion of additional energy sources, such as recombination energy, which may effectively increase the ejection efficiency \citep[see][for a review]{Ivanova2013}.

An alternative scenario is that the system formed through stable, non-conservative mass transfer from an evolved, giant-branch donor star. In this pathway, the binary would not have undergone a CE phase. Instead, the donor star began transferring mass to the white dwarf as it ascended the giant branch. For the system to remain stable and avoid a delayed dynamical-timescale mass transfer event, the mass ratio must have been such that the donor was less massive than the white dwarf, or the mass transfer must have been highly non-conservative.

While this scenario could produce wide-orbit systems, it is generally considered a less probable evolutionary channel for creating IPs. The expected mass transfer rates from giant donors are typically high, which should lead to rapid spin-down of the white dwarf via the "accretion-jet" torque or bury its magnetic field entirely, preventing it from manifesting as an IP \citep[e.g.][]{Sokoloski2004,Cumming2002}. If the \nsco\ 77.9~s  period in \nsco\ is rotational, it must have experienced a prolonged phase of spin-up, which is more characteristic of accretion from a main-sequence or sub-giant donor in a post-CE system.

\nsco\ is not entirely unique, but it may belong to the small group of known, long-period IPs. Systems such as AE~Aqr, V2731~Oph and Swift~J2006.4+3645 all exhibit orbital periods above the gap combined with rapid spin periods. These "wide-orbit IPs" challenge the paradigm that magnetic braking is the dominant mechanism for driving angular momentum loss and mass transfer in CVs, as it is thought to become inefficient for periods longer than $\sim$6-7 hours \citep{Verbunt1981}. In these wide systems, gravitational radiation is far too weak to drive the observed accretion rates. Therefore, they must either be in a post-nova phase of enhanced mass transfer, or the donor star itself must be undergoing nuclear evolution, driving mass loss through its own expansion. The existence of \nsco\ strengthens the evidence for a distinct population of IPs formed through alternative evolutionary channels. To distinguish between the inefficient CE and evolved-donor scenarios, spectroscopic observations to determine the donor star's spectral type, mass, and radius are essential. If the donor is found to be a main-sequence star, it would strongly favor the inefficient CE model and provide a crucial data point for calibrating binary evolution codes.

\subsection{\nher\ -- A High-Cadence View of Post-Nova Recovery}

Our re-examination of \tess\ observations for \nher\ confirms its nature as an IP, with a spin period of \mbox{501.328-s} and orbital period of \mbox{0.1529-days}. While these findings are consistent with previous work \citep{2022ApJ...940L..56P,Schaefer2022,2023MNRAS.521.5453S}, the high-cadence, uninterrupted \tess\ photometry allows us to probe the system's recovery in the aftermath of its 2021 nova outburst. The most significant insight from this analysis is the timing of the orbital signal's re-emergence.

A nova eruption is a cataclysmic event that is expected to completely disrupt, if not destroy, a pre-existing accretion disk. The intense radiation and expanding shell of ejecta should heat, ionize, and viscously strip away the disk material \citep{Livio1990}. A key question is how quickly the accretion disk can re-establish itself, allowing accretion to resume and the system to return to its quiescent state.

Our detection of a coherent orbital modulation just 17 days after the peak of the outburst, confirming results from \cite{2022ApJ...940L..56P} and \cite{2024BAAA...65...60L}, and the detection of the spin period even as early as 5 days after outburst by \cite{2022ApJ...940L..56P} can be interpreted as evidence that accretion restarted. The 17-day timescale is remarkably short when compared with, for example, \nsco\ on which the orbital period was not detected in the \tess\ observations obtained 43 days after outburst (see Fig. \ref{fig:psdV1716}). This implies that if a disk existed, either a substantial part of it survived the initial blast, perhaps shielded by its own bulk, or that the process of disk rebuilding from the incoming stream is far more rapid than many models have predicted. This finding provides a valuable benchmark for hydrodynamical simulations of post-nova accretion flows, suggesting that the viscosity and angular momentum transport in the newly-forming disk are highly efficient. Figure \ref{fig:psdV1674} shows a profile dominated by a single-humped sinusoid, which can also be interpreted as due to the irradiation of the companion star by its white dwarf companion. The lack of flux calibration and sector-to-sector varying background precludes to draw further conclusions about possible changes in profile amplitude linked to a physical origin such as changes in the illuminating source.

\section{Conclusions \label{sec:conc}}

Our analysis of \tess\ light curves from three novae uncovered unknown periodicities that suggest or confirm their magnetic nature. We summarize our findings and conclusions. 

\begin{itemize}

    \item In the 2024 \tess\ observations of \ncas, we found a period of approximately 116.88 seconds most likely related with the spin period of a strongly magnetized white dwarf. The period is rapidly evolving towards greater values (spin down), at one of the fastest rates recorded of \mbox{$\dot{P}=0.00165$ s d$^{-1}$} \citep[see e.g.][]{bruch_2025}. The detection of both beat periods ($\omega - \Omega$ and $\omega + \Omega$) in the power spectra points to a complex accretion geometry, likely dominated by a hybrid disk/stream-fed process. The presence of the $\omega + \Omega$ sideband, in particular, implies that a direct stream impact on the magnetosphere is occurring, providing a highly efficient mechanism to extract angular momentum and drive the extreme spin-down we observe.

    \item A comparison with the archetypal propeller AR~Sco reveals that \ncas, while sharing similar spin and orbital periods, is over four orders of magnitude more luminous and is spinning down five orders of magnitude faster, suggesting it represents a nascent, highly energetic stage of magnetic propeller evolution.
    
   \item We report the finding of a \mbox{1.35715$\pm$0.00487 d} period that is most likely related to the orbital motion in \nsco. This period is significantly detected in quiescence both before and after the recent nova outburst. In turn, this period is not detected during the fading phase after the outburst that was covered with \tess\ Sector 66. This is likely because the optical light during that phase was still dominated by the optically thick nova ejecta, which veiled the modulation from the underlying binary system, or accretion did not resume until later.

   \item A re-examination of \tess\ observations of \nher\ shows that the orbital period can be detected 17 days after the outburst. The recent observations during Sector 80 show a strong signal at the frequency corresponding to the spin period of \mbox{501.328$\pm$0.024-s}.  

\end{itemize}

\begin{figure}
\begin{center}
\includegraphics[scale=0.5]{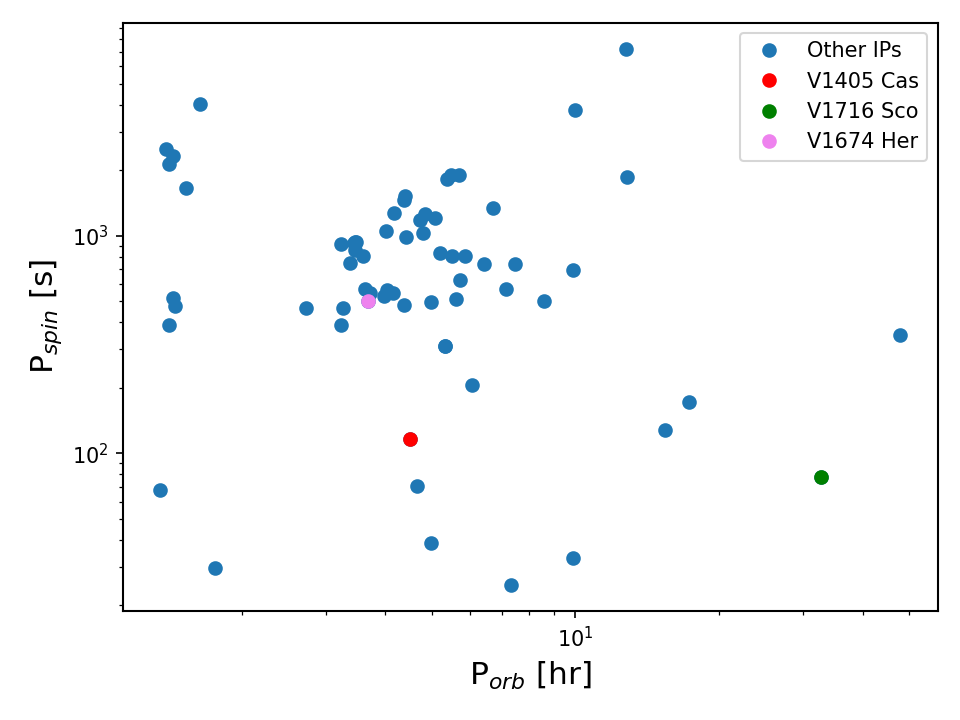}
\caption{The spin period versus orbital period diagram for magnetic cataclysmic variables. The population of known intermediate polars (IPs, blue circles) together with the three novae analyzed in this work being highlighted: \ncas\ (red circle), \nsco\ (green circle), and \nher\ (magenta circle). }
\label{fig:pspin_porb}
\end{center}
\end{figure}

Figure \ref{fig:pspin_porb} provides a crucial contextualization of our results by plotting the spin period as a function of the orbital period for the IPs. Our targets \nher\ and \ncas\ are located well within the dense cluster of known IPs, reinforcing our proposed classification within this group. The position of \ncas\ on this diagram, however, is particularly revealing. Plotted at an orbital period of $\sim$4.52 hours and a spin period of 116.88 seconds, it falls within the IP region but its pronounced spin-down distinguishes it from its neighbors. This rapid braking is causing the white dwarf to evolve rapidly in this parameter space; specifically, it is moving vertically upwards on this plot as its spin period increases. Its current state as a magnetic propeller, where accretion-powered luminosity is suppressed in favor of spin-down power, marks it as a system in a distinct evolutionary phase.

\begin{acknowledgements}

We acknowledge the anonymous referee for their careful reading and comments that significantly improved the manuscript. GJML is member of the CIC-CONICET (Argentina). We acknowledge Kim Page for the comments that considerably improved the manuscript. This work has made use of data from the All-Sky Automated Survey for Supernovae (ASAS-SN), as described by Shappee et al. (2014) and Kochanek et al. (2017). We also utilized data from the ASAS-SN Sky Patrol (Jayasinghe et al. 2023). 

\end{acknowledgements}

%
%

\bibliographystyle{aa}
\bibliography{listaref_MASTER}

\end{document}